\documentclass[reprint,pra,superscriptaddress,nofootinbib,aps]{revtex4-2}
\usepackage{amsmath, amssymb}	
\usepackage{bbm}				
\usepackage{braket}				
\usepackage{graphicx} 			
\usepackage{mathtools}			
\usepackage{epstopdf}
\usepackage{hyperref}			
\usepackage{cleveref}
\usepackage{tikz}
\usepackage{dsfont}
\usepackage{bm}
\usepackage[T1]{fontenc}
\usepackage{mathrsfs}
\usepackage{algorithm}
\usepackage{algpseudocode}
\allowdisplaybreaks[2]
\usetikzlibrary{shapes,snakes}
\usetikzlibrary{decorations.pathmorphing}
\tikzset{snake it/.style={decorate, decoration=snake}}
\algrenewcommand\algorithmicrequire{\textbf{Input:}}
\algrenewcommand\algorithmicensure{\textbf{Output:}}

\begin{document}

\title{Multiple uncertainty relation for accelerated quantum information}

\author{Chen Qian}
\email[]{qianch18@mail.ustc.edu.cn}
\affiliation{%
Shanghai Branch, National Laboratory for Physical Sciences at Microscale, University of Science and Technology of China, Shanghai 201315, China}
\affiliation{Shanghai Branch, CAS Center for Excellence and Synergetic Innovation Center in Quantum Information and Quantum Physics, University of Science and Technology of China, Shanghai 201315, China}
\affiliation{Department of Modern Physics, University of Science and Technology of China, Hefei, 230026, China}
\author{Ya-Dong Wu}
\email[]{yadongwu@hku.hk}
\affiliation{QICI Quantum Information and Computation Initiative, Department of Computer Science, The University of Hong Kong, Pok Fu Lam Road, Hong Kong}
\affiliation{Institute for Quantum Science and Technology, University of Calgary, Calgary, Alberta, T2N 1N4, Canada}
\affiliation{%
Shanghai Branch, National Laboratory for Physical Sciences at Microscale, University of Science and Technology of China, Shanghai 201315, China}
\author{Jia-Wei Ji}
\email[]{jiawei.ji@ucalgary.ca}
\affiliation{Institute for Quantum Science and Technology, University of Calgary, Calgary, Alberta, T2N 1N4, Canada}
\author{Yunlong Xiao}
\email[]{mathxiao123@gmail.com}
\affiliation{Nanyang Quantum Hub,  School of Physical and Mathematical Sciences, Nanyang Technological University, Singapore 637371, Singapore}
\affiliation{Complexity Institute, Nanyang Technological University, Singapore 639673, Singapore}
\affiliation{Department of Mathematics and Statistics, University of Calgary, Calgary, Alberta T2N 1N4, Canada}
\affiliation{Institute for Quantum Science and Technology, University of Calgary, Calgary, Alberta, T2N 1N4, Canada}
\author{Barry C. Sanders}
\email[Corresponding author, ]{bsanders@ustc.edu.cn}
\affiliation{%
Shanghai Branch, National Laboratory for Physical Sciences at Microscale, University of Science and Technology of China, Shanghai 201315, China}\affiliation{Shanghai Branch, CAS Center for Excellence and Synergetic Innovation Center in Quantum Information and Quantum Physics, University of Science and Technology of China, Shanghai 201315, China}
\affiliation{Institute for Quantum Science and Technology, University of Calgary, Calgary, Alberta, T2N 1N4, Canada}
\begin{abstract}
The uncertainty principle,
first introduced by Heisenberg in inertial frames, clearly distinguishes quantum theories from classical mechanics. In noninertial frames, its information-theoretic expressions, namely entropic uncertainty relations, have been extensively studied through delocalized quantum fields, and localization of the quantum fields were discussed as well.
However, infeasibility of measurements applied on a delocalized quantum field due to the finite size of measurement apparatuses is left unexplained.
Therefore, physical clarification of a quantum protocol revealing entropic uncertainty relations still needs investigation.
Building on advances in quantum field theories and theoretical developments in entropic uncertainty relations, we demonstrate a relativistic protocol of an uncertainty game in the presence of localized fermionic quantum fields inside cavities. 
Moreover, a novel lower bound for entropic uncertainty relations with multiple quantum memories is given in terms of the Holevo quantity, which implies how acceleration affects uncertainty relations.
\end{abstract}
\maketitle
\section{Introduction}
Quantum mechanics offers us a new way to transform information, i.e., encoding information into a state of a quantum system and decoding information through quantum measurements, which raises the curtain of quantum information theory.
Besides the advantages provided by quantum mechanics,
quantum mechanics also imposes strict restrictions on what we can gain from quantum measurements. These restrictions are known as the Heisenberg uncertainty principle, which lies at the heart of quantum theory.
A decade ago, entropic uncertainty relations including quantum memories,
which allows entanglement with a measured system,
were introduced~\cite{nature-2010} and paved the way to applications
involving entanglement witnesses and quantum key distribution~\cite{rmp-2017}.
In particular,
the entropic uncertainty relations of tripartite systems quantify the information tradeoff for two incompatible
observables stored in two separate quantum memories~\cite{prl-2009}
and reveals monogamy of entanglement,
which lies at the center of the debate of fundamental puzzles, such as the black-hole firewall paradox~\cite{firewall-jhep-2013}.

Relativistic quantum information is a fast-growing field that hopefully could explain these puzzles~\cite{prl-2005,pra-2006,Martinez-PhD,cqg-2012-Martinez,cqg-2012-Alsing,cqg-2014}.
Relativistic effects on entropic uncertainty relations with
quantum memories have been extensively studied for bipartite systems~\cite{plb-2013-fan1, plb-2015-fan2, laserpl-2017, andp-2018, epjc-2018},
but some important aspects are infeasible.
On the one hand, previous works on uncertainty relations under relativistic effects are based on delocalized quantum fields, whose excitations could thus be created anywhere in the whole Universe.
Regarding this point,
applying operations on quantum information over all spacetime for a pointlike observer is unexplained.
On the other hand, the acceleration horizon under the Unruh effect
exists only when an observer accelerates from the asymptotic past to the asymptotic future~\cite{unruh-1976}.
In this setting, how to describe the observer's motion if the observer only accelerates until the measurements are performed has not been made clear.
Despite Feng \textit{et al.}~having mentioned the localization of quantum fields in bipartite entropic uncertainty relations~\cite{plb-2013-fan1}, they focused only on calculations of the uncertainty bound after acceleration, and left a detailed physical description on the relativistic protocol of the uncertainty game unexplained. 
Therefore, in order to analyze the relativistic effects on uncertainty relations clearly,
we need to localize quantum-field information and clarify the uncertainty game in a physical scenario.

In addition to these gaps in knowledge,
multipartite uncertainty relations for noninertial frames have not been investigated yet.
Previous lower bounds for entropic uncertainty relations for tripartite systems are state independent~\cite{prl-2009},
which do not depend on relativistic motion of quantum memories.
For puzzles like the black-hole paradox,
relativistic effects on bipartite uncertainty relations must be generalized to the case of multipartite systems.

Therefore, our aim in this paper is to investigate multipartite entropic uncertainty relations for localized quantum information in noninertial frames.
Here, we focus on fermionic quantum fields localized in rigid cavities, which can be accelerated. The length of a rigid cavity is immune to relativistic effects during acceleration, due to different accelerations of two walls of the cavity~\cite{Friis-PhD}.
For simplicity, we consider spinless (spin-polarized) fermions confined in a one-dimensional optical lattice inside cavities. This can be accomplished by cooling unpolarized interacting fermions in optical lattices across different cavities, where they spontaneously form a Mott insulating state (i.e., with one fermion per cavity) characterized by antiferromagnetic spin correlations~\cite{Mazurenko2017,Boll1257,Parsons1253}. Then, we select one spin component by applying a global resonant laser pulse~\cite{Parsons1253,Brown1385} and an external magnetic field. Moreover, in this work we wish to use one spinless fermion to encode quantum information across three cavities, which requires us to remove other fermions and allow it to hop between cavities by controlling the lattice potential.

The model of an accelerating cavity has been used to study relativistic effects on quantum entanglement~\cite{bos-prd-2012,fer-prd-2012,Friis-PhD}, quantum teleportation~\cite{Friis-prl-2013}, quantum secret sharing~\cite{Ahmadi-prd-2017}, and quantum clock~\cite{Lindkvist-pra-2014}. In the following, we combine the transformation of localized fermionic fields in noninertial frames with
multipartite entropic uncertainty relations to formulate a relativistic protocol.

Our protocol is depicted in Fig.~\ref{cavity}, where the cavities,
serving as quantum memories, go through relativistic motions~\cite{prl-2009}. We offer an irreducible bound of entropic uncertainty in terms of Holevo quantities for multipartite systems
for quantum information stored in accelerating quantum memories.
The difference between uncertainty and its lower bound is independent of the acceleration of quantum memories, whereas the lower bound itself changes with acceleration.
We find that,
as entanglement between Alice and Bob increases,
Bob has a higher possibility of guessing Alice's outcome correctly when Alice measures an arbitrary observable $M_1$,
whereas the possibility for Charlie to have a good guess decreases when Alice measures another incompatible observable $M_2$,
yielding a nonzero lower bound for the uncertainty relations.

In what follows, we first briefly the review background for entropic uncertainty relations with quantum memories and the evolution of fermionic quantum fields inside accelerating cavities in Sec.~\ref{sec_back}, and then give a detailed introduction to our relativistic protocol and new lower bound in Sec.~\ref{sec_approach}. We present a specific case of the protocol and show our calculation results in Sec.~\ref{sec_results}. Finally, we discuss our results and provide future lines of research in Sec.~\ref{sec_dis}. The proofs of our new bound are given in the Appendixes. Throughout this paper, we use units in which $\text{c} = \hbar = 1$ and base-2 logarithms.

\section{\label{sec_back}Preliminaries}

In this section we give a review on entropic uncertainty relations, especially the results under relativistic effects, after which
we review the physical model used in this paper, namely, a fermionic field inside an accelerating cavity.

\subsection{Entropic uncertainty relations in relativistic systems}

In this subsection, we provide a brief review on entropic uncertainty relations with quantum memories and its development in relativistic systems.
An entropic uncertainty relation with quantum memories was first introduced for entangled bipartite systems~\cite{nature-2010}.

Suppose there are three agents, Alice, Bob, and a dealer. The dealer prepares an entangled bipartite state $\rho_\text{AB}$, and sends subsystem~A to Alice and subsystem~B to Bob.
Subsystem~B is regarded as a quantum memory of subsystem~A, which means that Bob can access information from Alice's subsystem. Alice randomly chooses
to measure either observable $M_1$ or observable $M_2$ on the $n$-dimensional Hilbert space $\mathscr{H}_\text{A}$. While broadcasting her measurement choice to Bob,
Alice keeps the measurement outcomes confidential. Bob's uncertainty about Alice's measurement outcomes can be expressed as the summation of conditional entropies
\begin{equation}
\label{eq:measure-cond}
	H(M_i|\text{B}):=H(\rho_{M_i\text{B}})-H(\rho_\text{B}), i\in\{1,2\},
\end{equation}
where 
\begin{equation}
\label{eq:Shannon-rho}
H(\rho)
		=-\sum_{i=1}^n p_i \log p_i
\end{equation}
denotes the Shannon entropy of an $n\times n$ density matrix~$\rho$ with $\{p_i\}$ eigenvalues of $\rho$, it is equivalent to the von Neumann entropy $S(\rho)=-\text{Tr}(\rho\log \rho)$, and
\begin{equation}
\label{eq:post-measure}
\rho_{M_{i}\text{B}}:=\sum^n_{j=1} (\ket{u_j^i}\bra{u_j^i} \otimes \mathds{1}) \rho_{\text{AB}} (\ket{u_j^i}\bra{u_j^i} \otimes \mathds{1})
\end{equation}
is the postmeasurement state after measuring $M_{i}$ on Alice's system,
and $\ket{u_j^i}$ denotes the $j$th eigenvector of $M_i$.

With these notations, we are now in the position to formulate an uncertainty relation in the presence of quantum memory, namely,
\begin{equation}
\label{eq:bimem}
	H(M_1|\text{B})+H(M_2|\text{B})\geq \log\frac{1}{c}+H(\text{A}|\text{B}),
\end{equation}
with
\begin{equation}
c=\max_{i,j}|\braket{u_i^1|u_j^2}|^2,
\end{equation}
denoting the maximal overlap between the eigenvectors of $M_1$ and $M_2$, and
\begin{equation}
\label{eq:condent}
	H(\text{A}|\text{B}):=H(\rho_\text{AB})-H(\rho_\text{B})
\end{equation}
is the conditional entropy quantifying the side information~B possesses about~A.
As a consequence of Eq.(\ref{eq:bimem}), we know that the bipartite entanglement shared between Alice and Bob can reduce Bob's uncertainty about Alice's measurement outcomes.

To further investigate the interconnections between the entanglement theory and the uncertainty principle, let us consider the case with entanglement distributed in a tripartite system. More precisely, we consider four agents: Alice, Bob, Charlie, and a dealer. The dealer prepares a state $\rho_{\text{ABC}}$ and distributes the subsystem~A to Alice, the subsystem~B to Bob, and the subsystem~C to Charlie. Alice measures either $M_1$ or $M_2$ at the subsystem~A and broadcasts her choice to Bob and Charlie.
The ability of Bob and Charlie to infer Alice's measurement outcomes correctly is limited by~\cite{nature-2010,prl-2012}
\begin{equation}
\label{trimem}
H(M_1|\text{B})+H(M_2|\text{C})\geq\log\frac{1}{c}.
\end{equation}
Remarkably, the bound~(\ref{trimem})
 expresses not only the uncertainty principle, but also the monogamy of entanglement~\cite{rmp-2017}.

In recent years, entropic uncertainty relations are studied in relativistic systems with delocalized fields~\cite{plb-2013-fan1, plb-2015-fan2, laserpl-2017, andp-2018, epjc-2018}.
According to this research, bipartite entropic uncertainty relations 
depend on the acceleration of a quantum memory in noninertial frames due to the Unruh effect.
When acceleration increases, uncertainty increases due to degradation of entanglement.
The result is applicable to the uncertainty relations for Hawking radiation of a Schwarzschild black hole~\cite{plb-2015-fan2, andp-2018, epjc-2018}.

A new lower bound was derived for a bipartite entropic uncertainty relation in the presence of quantum memory via the Holevo quantity~\cite{epjc-2018}. This new lower bound is advantageous,
because as the memory accelerates,
the difference between the uncertainty and its lower bound remains independent of acceleration.
Their uncertainty relation is
\begin{align}
\label{binew}
	H(M_1|\text{B})&+H(M_2|\text{B})\nonumber\\
		&\geq\log\frac{1}{c}+H(\text{A})-\mathcal{J}(\text{B}|M_1)
			-\mathcal{J}(\text{B}|M_2),
\end{align}
where
\begin{align}
\label{Holevo}
&\mathcal{J}(\text{B}|M_i)=H(\rho_\text{B})-\sum_j^n p^i_j H(\rho_{\text{B}|u^i_j}) ,\\
&\rho_{\text{B}|u^i_j}=\bra{u^i_j}\rho_\text{AB}\ket{u^i_j}/\text{Tr}(\bra{u^i_j}\rho_\text{AB}\ket{u^i_j}), i\in\{1,2\},
\end{align}
 and $p^i_j$ is the $j$th eigenvalue of $M_i$, respectively.
$\mathcal{J}(\text{B}|M_i)$ is the Holevo quantity, which reveals how much information is accessible from quantum memory~B about measurement outcomes of $M_i$.

\subsection{Relativistic quantum information in cavities}

After introducing entropic uncertainty relations and their applications to relativistic systems,
we briefly review the effect of acceleration on a fermionic field inside a cavity,
which has been well studied during the past decade~\cite{fer-prd-2012,bos-prd-2012,Friis-PhD}.

\begin{figure}
\centering
\includegraphics[width=0.5\textwidth]{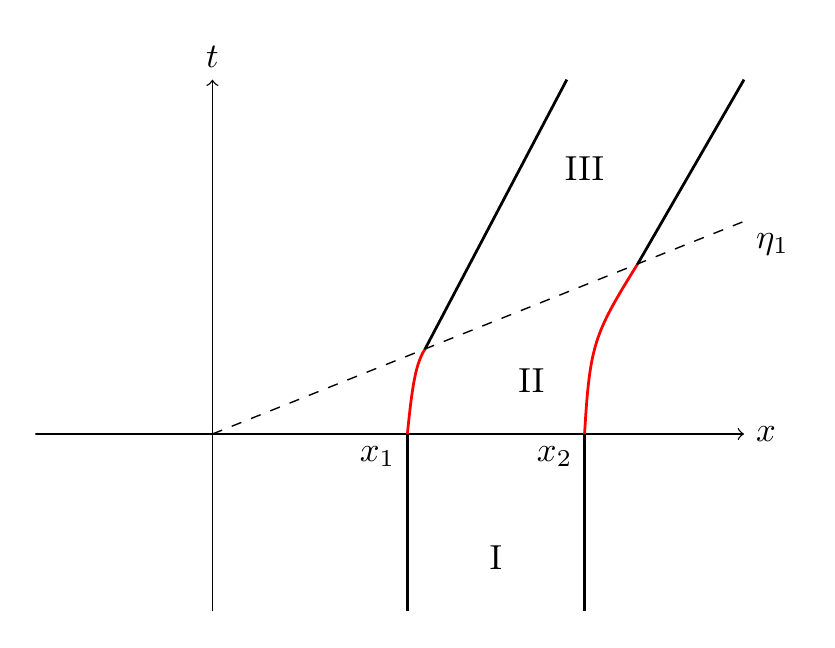}
\caption{\footnotesize The thick lines show a BBB trajectory of a cavity with the proper length $x_2-x_1$, where $x_1$ and $x_2$ are the spatial coordinates
of the left and right walls of the cavity at time $t=0$, respectively. The thick lines are world lines of the left and right walls of the cavity are depicted; they include three parts. 
At Rindler time $\eta_1$, the cavity stops accelerating and returns to an inertial frame.}
\label{BBB}
\end{figure}

We focus on the basic building block (BBB) trajectory of a rigid cavity experiencing any arbitrary nonuniform motion~\cite{Friis-PhD}. As shown in Fig.~\ref{BBB}, the whole trajectory comprises three steps.
Initially, the cavity is at rest in region~I.
After $t= 0$, the cavity accelerates with a constant acceleration in region~II.
The proper acceleration at the center of the cavity is $a=\frac{2}{x_1+x_2}$.
In region~III, the cavity stops accelerating and moves uniformly. The BBB starts from an inertial cavity and ends with another inertial cavity, with a single intermediate period of uniform acceleration.
The transition of a fermionic quantum field inside the cavity from the inertial region to the uniformly accelerating region can be represented as a linear transformation of the modes, which is known as the Bogoliubov transformation~\cite{Friis-PhD}.

Now we illustrate the Bogoliubov transformation in detail. In our protocol, the quantum field inside the cavity is a spinless fermionic field, so we focus on the transformation of fermionic fields. Modes of the field~$\{\psi^{\text{I}}_k\}$, in region~I, are classified by positive and negative frequencies with respect to $\partial_t$ (the future-directed Minkowski Killing vector)~\cite{pra-2006}.

In region~I, the field can be expanded as
\begin{equation}
	\psi^{\text{I}}=\sum_{k\ge 0}\hat{a}_k\psi^{\text{I}}_k+\sum_{k<0}\hat{b}^\dagger_k\psi^{\text{I}}_k,
\end{equation}
where $\hat{a}_k$ is the annihilation operator of a positive mode and $\hat{b}^\dagger_k$ is the creation operator of a negative mode, and $k\in \mathbb N$ is a mode label.
Different modes are orthonormal~\cite{pra-2006}
\begin{equation}
\label{orthonormal-relation}
	(\psi^{\text{I}}_{k'},\psi^{\text{I}}_ k)=\delta_{k'k},
\end{equation}
where
\begin{equation}
\label{dirac-product}
	\left(\psi_{k'}, \psi_{k}\right)=\int^{x_2}_{x_1} \text{d}x \; \psi^{\dag}_{k'}(x,t)  \psi_{k}(x,t)
\end{equation}
denotes the Dirac inner product.
The creation and annihilation operators satisfy anticommutation relations
\begin{equation}
    \left\{\hat{a}_{k'},\hat{a}^\dagger_ k\right\}=\left\{\hat{b}_{k'},\hat{b}_k^\dagger\right\}=\delta_{k'k}\mathds1.
\end{equation}
$\{\psi^{\text{I}}_k\}$ forms a complete set of orthogonal mode functions of the fermionic field in region~I.

When the field transforms from region~I to region~II at $t=0$, the set of mode functions $\{\psi^{\text{I}}_k\}$ transforms to $\{\psi^{\text{II}}_k\}$. Then we have the Bogoliubov transformation from region~I to region~II,
namely
\begin{equation}
	\psi^{\text{II}}_{k'}=\sum_{k} F_{k' k}\psi^{\text{I}}_k  ,  
\end{equation}
where
\begin{equation}
	F_{k'k} :=\left(\psi^{\text{II}}_{k'}, \psi^{\text{I}}_k\right)
\end{equation}
is called the Bogoliubov coefficient, and the Bogoliubov transformation matrix $F$ is a unitary matrix.

In our work, $F_{k'k}$ is expanded in terms of the perturbative variable
\begin{equation}
    h:=aL,
\end{equation}
where
\begin{equation}
\label{eq:cavitylength}
	L=x_2-x_1
\end{equation}
is the proper length of the cavity. Up to $h^2$, we have
\begin{equation}
\label{eq:Fk'k}
    F_{k'k}
    	=F_{k'k}^{(0)}+F_{k'k}^{(1)}h+F_{k'k}^{(2)}h^2+\mathcal{O}(h^3).
\end{equation}
Perturbation elements $F_{k'k}^{(i)}$ in~(\ref{eq:Fk'k}) have been calculated in detail~\cite{fer-prd-2012}.

As motions of the cavities in regions~I and~III are related by a Lorentz transformation, the transformation between regions~II and~III is simply the inverse transformation $F^{-1}$. We choose the same phase conventions for region~III solutions
at $\eta=\eta_1$ as region~I solutions at $\eta=0$ (we set the phase at $\eta=0$ to be $0$). The phases acquired by the Rindler modes in region~II are accounted for
by a diagonal matrix $G(\eta_1)$,
where the $n \text{th}$ diagonal entry of the matrix is
\begin{equation}
G_n(\eta_1)=\exp{(i\Omega_n \eta_1)}=\text{exp}\frac{i\pi(n+s)\eta_1}{\ln{(x_2/x_1)}},
\end{equation}
where $s \in [0,1)$ is a parameter characterizing the phase shifts of reflections from the two walls of the cavity.

Overall, the Bogoliubov transformation from region~I to region~III is
\begin{equation}
	\mathbb{F}=F^{-1}G(\eta_1)F=F^\dagger G(\eta_1)F.
\label{eq:I to III bogoliubov}
\end{equation}
Therefore, the transformation of fermionic fields from region~I to region~III is
\begin{equation}
\psi^{\text{III}}_{k'}=\sum_{k} \mathbb{F}_{k'k}\psi^{\text{I}}_k  , 
\end{equation}
where $\mathbb{F}_{k'k}$ can also be expanded perturbatively in $h$ as
\begin{equation}
\mathbb{F}_{k'k}
	= \mathbb{F}_{k'k}^{(0)}+\mathbb{F}_{k'k}^{(1)}h+\mathbb{F}_{k'k}^{(2)}h^2+\mathcal{O}(h^3).
\end{equation}
with
\begin{subequations}\label{bogo-coeff}
\begin{align}
& \mathbb{F}_{k'k}^{(0)}= \delta_{k'k}G_k,
\\[1ex]
& \mathbb{F}_{k'k}^{(1)}
	= (G_{k'}-G^*_{k}) F_{k'k}^{(1)},
\\[1ex]
& \mathbb{F}_{kk}^{(2)}= G_{k}F_{kk}^{(2)}+F_{kk}^{*(2)}G_{k}+\sum_{k'}F_{k'k}^{* (1)}G_{k'}F_{k'k}^{(1)}.
\end{align}
\end{subequations}
by using Eq.~(\ref{eq:I to III bogoliubov}) and Eq.~(\ref{eq:Fk'k}). Since our calculation below does not include $\mathbb{F}_{k' k}^{(2)}$ ($k' \ne k$), we just show the expansion of $\mathbb{F}_{kk}^{(2)}$ here.

For specific calculations to come,
we consider the Bogoliubov transformation of the vacuum and one-particle states inside the cavity from region~I to region~III, which was used to study how entanglement is affected due to the relativistic motion with quantum information encoded in fermionic modes~\cite{fer-prd-2012,Friis-PhD}. First, we assume that all the modes chosen in our calculations are positive modes. Vacuum state in region~I is denoted by $\ket0^{\text{I}}$; then the one-particle state is $\ket{1_k}^{\text{I}}=\hat{a}^\dagger_k\ket0^{\text{I}}$. When the cavity stops accelerating at $\eta=\eta_1$ and moves uniformly in region~III, we denote vacuum state of the field inside it by $\ket0^{\text{III}}$, and the one-particle state is $\ket{1_k}^{\text{III}}=\hat{d}^\dagger_k\ket0^{\text{III}}$, where $\hat{d}^\dagger_k$ is the creation operator in region~III.

Suppose $k\ge 0$ and all the modes except~$k$ are in vacuum states initially. By tracing out all the modes except~$k$~(denoted by $\lnot k$), we get the transformation of the vacuum and one-particle state of the field from region~I to region~III as~\cite{fer-prd-2012}
\begin{subequations}
\begin{align}
& \text{Tr}_{\lnot k}\ket0^{\text{I}}\bra0 =\left(1-f^-\right)\ket{0_k}^{\text{III}}\bra{0_k}+f^-\ket{1_k}^{\text{III}}\bra{1_k},
\\[1ex]
& \text{Tr}_{\lnot k}\ket0^{\text{I}}\bra{1_k}=\left(G_k+\mathbb{F}_{kk}^{(2)}h^2\right)\ket{0_k}^{\text{III}}\bra{1_k},
\\[1ex]
& \text{Tr}_{\lnot k}\ket{1_k}^{\text{I}}\bra{1_k}=(1-f^+)\ket{1_k}^{\text{III}}\bra{1_k}+f^+\ket{0_k}^{\text{III}}\bra{0_k},
\end{align}
\label{trans from I to III}
\end{subequations}
where
\begin{subequations}
\begin{align}
&f^+= \sum_{l \geq 0}\left|\mathbb{F}^{(1)}_{lk}\right|^2=\sum_{l \geq 0}|\text{e}^{2\pi \text{i} u(k-l)}-1|^2\left|F_{kl}^{(1)}h\right|^2,\\
&f^-= \sum_{l < 0}\left|\mathbb{F}^{(1)}_{lk}h\right|^2,
\end{align}
\end{subequations}
and
\begin{equation}\label{expressionofu}
	u=\frac{\eta_1}{2\ln{(x_2}/{x_1})}=\frac{\eta_1}{2\ln [(a\cdot L+2)/(2-a\cdot L)]}
\end{equation}
is a factor related to the acceleration of the cavity.

\section{\label{sec_approach}Approach}
\begin{figure}
\centering
\includegraphics[width=0.55\textwidth]{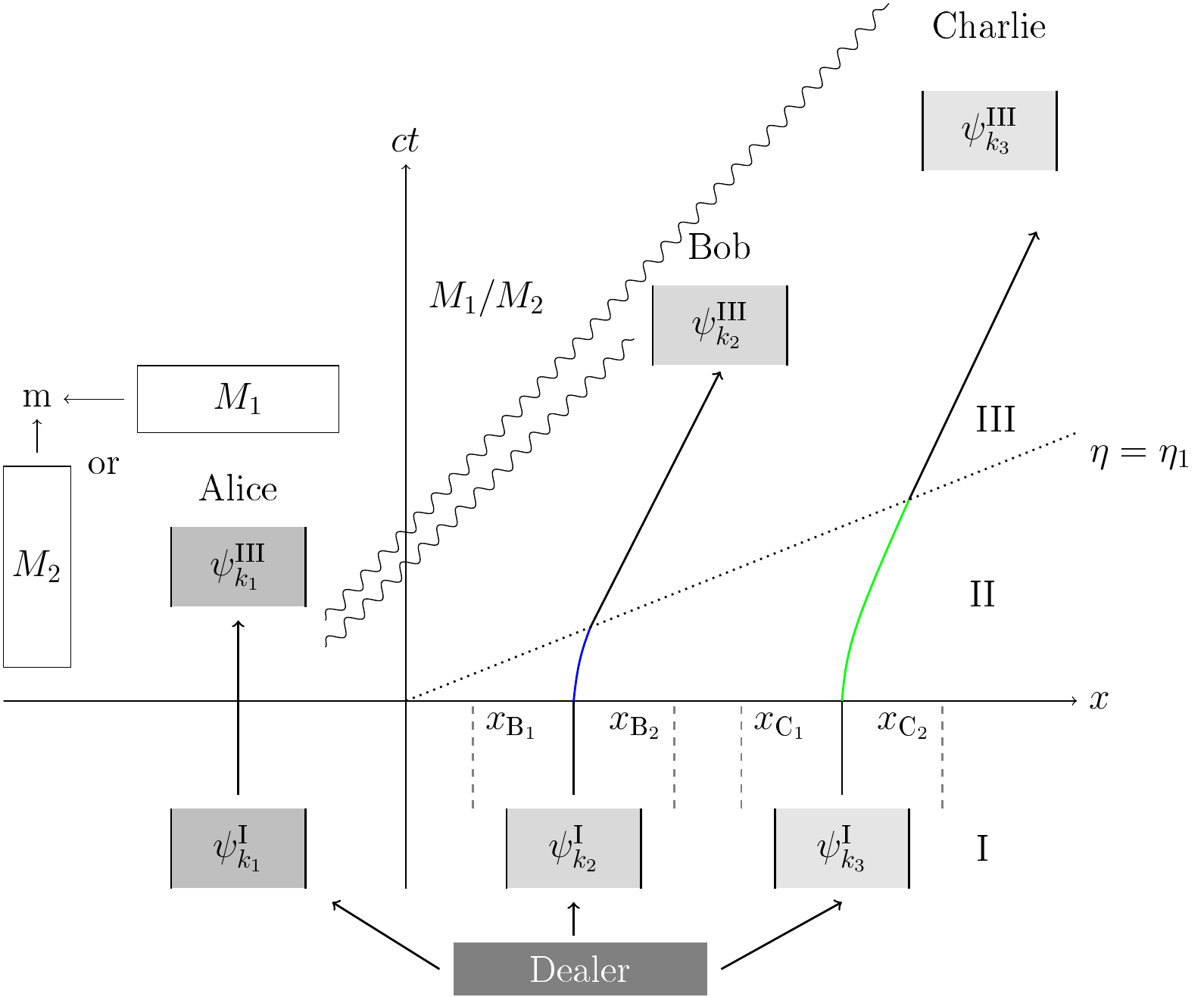}
\caption{\footnotesize The relativistic protocol with four agents, Alice, Bob, Charlie, and a dealer. $\psi^{\text{I}}_{k_1}$, $\psi^{\text{I}}_{k_2}$, and $\psi^{\text{I}}_{k_3}$ are the mode functions in the cavities of Alice, Bob, and Charlie, respectively, in region~I, and $\psi^{\text{III}}_{k_1}$, $\psi^{\text{III}}_{k_2}$, and $\psi^{\text{III}}_{k_3}$ are the mode functions in their cavities in region~III. $M_1$ and $M_2$ are observables Alice can randomly choose to measure at mode $\psi^{\text{III}}_{k_1}$ in her cavity, and $m$ is the outcome that she should keep secret. The waves are electromagnetic broadcastings of Alice's choice to Bob and Charlie, but Bob and Charlie are not allowed to communicate in region~III. The blue and green curves stand for the accelerating processes of Bob and Charlie's cavities in region~II before Rindler time $\eta_1$. Note that $x_{\text{B}_2}$$-$$x_{\text{B}_1}$ is the length of Bob's cavity and $x_{\text{C}_2}$$-$$x_{\text{C}_1}$ is the length of Charlie's cavity. }
\label{cavity}
\end{figure}

In this section, we develop a relativistic protocol for an uncertainty game by employing the evolution of fermionic fields inside the cavities moving with the trajectories of BBB in a protocol revealing the uncertainty relation in an entangled tripartite system, as depicted in Fig.~\ref{cavity}.

Four agents, Alice, Bob, Charlie and a dealer, enact our protocol.
At first, a dealer aligns three cavities with one spinless fermionic field inside each cavity and prepares a tripartite entangled state with one mode in each cavity. Then the dealer delivers the three cavities to Alice, Bob, and Charlie, respectively, and tells them their tasks, including how long Bob and Charlie should accelerate and when Alice should apply measurements. After $t= 0$, Bob and Charlie both accelerate, which is shown in region~II. Then they stop accelerating at Rindler time~$\eta_1$.

When they stop accelerating, Alice stays in her lab, and randomly chooses one of two detectors that can measure $M_1$ and $M_2$ respectively, to perform a measurement on the mode in her cavity. Then she broadcasts her choice of the measurement to Bob and Charlie, and keeps the measurement outcomes secret. After that, Bob and Charlie play a game against Alice without communication between themselves, where they win only if both of them guess the outcomes correctly. The game rule is that, Bob performs the measurement in his own cavity after Alice detects $M_1$, and Charlie performs the measurement in her own cavity after Alice detects $M_2$.
The protocol needs to be repeated for many rounds so that Bob and Charlie each get a distribution of their inferences.

The lower bound of the uncertainty relation is a limit of the ability for Bob and Charlie to both correctly infer Alice's outcomes. Here, we introduce a new lower bound of multiple uncertainty relations in terms of Holevo quantities by generalizing~(\ref{binew}).
Contrary to a previous state-independent bound like~($\ref{trimem}$), our bound reveals how acceleration affects the uncertainty. The difference between the joint uncertainty and our new bound is independent of the acceleration of memories.

Suppose $N+1$ physical systems share a multipartite entangled state. One agent applies the measurement and the other $N$ agents act as quantum memories $\text{E}_i$. The multiple uncertainty relation conditioned on quantum memories is
\begin{equation}
\label{Nmem}
	\sum_i^N H\left(M_i|\text{E}_i\right)\geq\mathcal{B}-\sum_i^N\mathcal{J}(\text{E}_i|M_i) ,
\end{equation}
where $\mathcal{J}(\text{E}_i|M_i)$ is the Holevo quantity~(\ref{Holevo}),
and~$\mathcal{B}$ is a state-independent lower bound,
which have been well-studied~\cite{pra-2015,pra-2016,pra-2018}.
Detailed derivations of Eq.~(\ref{Nmem}) are in Appendix~\ref{appendix_proof}.
Specifically, for a tripartite system as discussed in the uncertainty game we have introduced above, we choose $\mathcal{B}$ as $\log \frac{1}{c}$ and the uncertainty relation is
\begin{align}
\label{3mem}
	H\left(M_1|\text{B}\right)&+H\left(M_2|\text{C}\right)\nonumber\\
    &\geq \log \frac{1}{c}-\mathcal{J}(\text{B}|M_1)-\mathcal{J}(\text{C}|M_2).
\end{align}
According to the monogamy of tripartite entanglement, if Bob's inference about Alice's outcomes of $M_1$ is correct, then Charlie cannot infer Alice's outcomes of $M_2$ precisely.

\section{\label{sec_results}Results}
Suppose the dealer prepares a W state
\begin{align}
	\ket{\psi_0}_{\text{ABC}}
		=&\frac{1}{\sqrt{3}}\big(\ket{1_{k_1}0_{k_2}0_{k_3}}^{\text{I}}\nonumber\\
		&+\ket{0_{k_1}1_{k_2}0_{k_3}}^{\text{I}} +\ket{0_{k_1}0_{k_2}1_{k_3}}^{\text{I}}\big)
\label{W-state}
\end{align}
for Alice, Bob, and Charlie initially in region~I.
Then Alice stays at rest while Bob and Charlie move in the trajectories of BBB. From region~I to region~III, the modes entangled inside Alice, Bob, and Charlie's cavities experience a Bogoliubov transformation and the initial state~(\ref{W-state}) transforms
to a mixed state $\rho_{\text{ABC}}$.
We calculate $\rho_{\text{ABC}}$ by substituting~(\ref{trans from I to III}) into~(\ref{W-state}). Finally, when all of the three agents are in region~III, Alice chooses to measure either $M_1$ or $M_2$ on her mode inside her cavity, and broadcasts her choice to Bob and Charlie.

From the Jordan-Wigner transformation,
we know fermions can be mapped to a spin chain.
However, mappings between fermions and qubits are possible only for a few states due to superselection rules and other restrictions~\cite{pra-2013}.
We note that the W state is a special case, which does not violate these restrictions, as we show in Appendix~\ref{appendix_W}.
Therefore, we can consider fermionic modes in the W state as qubits, and set $M_1$ and $M_2$ as Pauli operators.

In general, due to the Jordan-Wigner transformation, Pauli operators are not local in a fermionic system. However, as only Alice's mode undergoes Pauli measurements, we can ignore the nonlocal phase here. The Pauli operators, in our paper, represent measurements of the superposition of particle numbers, where, if we measure $\sigma_z$, we can get $0$ for ``no particle'' and $1$ for ``one particle'', and if we measure $\sigma_x$ or $\sigma_y$, we can get the phase of the superposition of no particle and one particle.

When Alice applies a measurement of $\sigma_x$, $\rho_{\text{ABC}}$ changes to a postmeasurement state
\begin{equation}
\rho_{\sigma_x \text{BC}}=\sum_{i=+,-}(\Pi_i\otimes\mathds{1}\otimes\mathds{1})\rho_{\text{ABC}}(\Pi_i\otimes\mathds{1}\otimes\mathds{1}),
\end{equation}
where $\Pi_{\pm}=\ket{\pm}\bra{\pm}$ are projections onto the eigenvectors of $\sigma_x$ at subsystem $\text{A}$. Suppose Charlie guesses the outcome when Alice measures $\sigma_x$, and it is Bob's turn to guess when Alice measures $\sigma_y$. To calculate the conditional entropy $H(\sigma_x|\text{C})$, we express $\rho_{\sigma_x \text{C}}$ in the basis $\{\ket{00}, \ket{01}, \ket{10}, \ket{11}\}$ by tracing out the subsystem $\text{B}$ as
\begin{equation}
\label{sigxC}
{\rho}_{\sigma_x \text{C}}  = \frac{1}{6}
\begin{bmatrix}
2-\bar{f}_{k_3} & 0 & 0 & \mathcal{F}_{k_3}\\
0 & 1+\bar{f}_{k_3} & \mathcal{F}^*_{k_3} &0\\
0 & \mathcal{F}_{k_3} &2-\bar{f}_{k_3} & 0\\
\mathcal{F}^*_{k_3} & 0 & 0 & 1+\bar{f}_{k_3}  \\
\end{bmatrix}
,
\end{equation}
where
\begin{equation}
\label{parameter}
	\bar{f}_{k_3}=-f^{+}_{k_3}+2f^{-}_{k_3},\,
	\mathcal{F}_{k_3}=G_{k_3}+\mathbb{F}_{k_3k_3}^{(2)}h^2.
\end{equation}
$\rho_{\sigma_x \text{C}}$ has four eigenvalues
\begin{subequations}
\label{eigenfer-cav}
\begin{align}
& \lambda_{1}=\lambda_{2}= \frac{3-\sqrt{1-4\bar{f}_{k_3}+4\bar{f}_{k_3}^2+4\mathcal{F}_{k_3}^2}}{12},\\
& \lambda_{3}=\lambda_{4}= \frac{3+\sqrt{1-4\bar{f}_{k_3}+4\bar{f}_{k_3}^2+4\mathcal{F}_{k_3}^2}}{12}.
\end{align}
\end{subequations}
Then we calculate
\begin{align}
\label{Cavity-fermion-C}
H(\sigma_x|\text{C})=&H(\rho_{\sigma_x \text{C}})-H(\rho_{\text{C}}) \nonumber\\
	=&2H\left(\lambda_{1}\right)
		+ 2H\left(\lambda_{3}\right)\nonumber-H\left(\frac{2-\bar{f}_{k_3}}{3}\right)\\
		&-H\left(\frac{1+\bar{f}_{k_3}}{3}\right).
\end{align}
Similarly, we calculate $H(\sigma_y|\text{B})$ by obtaining the postmeasurement state $\rho_{\sigma_y \text{B}}$ using the same method as (\ref{sigxC}).

\begin{figure}
\centering
\includegraphics[scale=1.5]{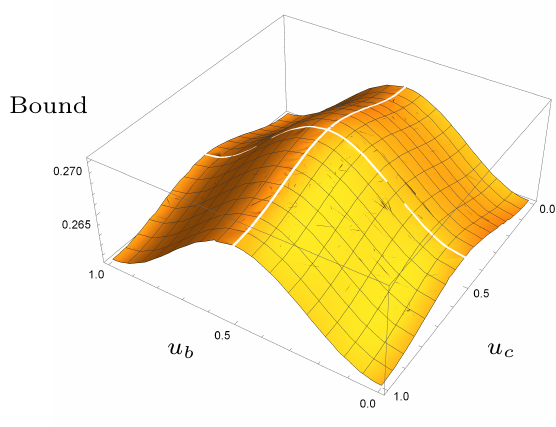};
\caption{\footnotesize The change of lower bound of the uncertainty relation with the lengths of cavities and accelerating time are fixed. $u_b$ and $u_c$ are acceleration factors for Bob's and Charlie's cavities, respectively. In this plot, we have set $h=0.1$ and chosen $k_2=2, k_3=1$, and for massless fermionic fields $s_2=s_3=\frac{1}{2}$~\cite{fer-prd-2012}.}
\label{result}
\end{figure}

After that, using relation (\ref{Nmemeq}) in Appendix~\ref{appendix_proof}, the lower bound of the uncertainty relation~(\ref{3mem}) can be evaluated by 
\begin{equation}
\label{bound}
	H(\sigma_x|\text{C})+H(\sigma_y|\text{B})-H(\sigma_x)-H(\sigma_y)+\log \frac{1}{c}\, .
\end{equation}
Our result is shown in Fig.~\ref{result}: the bound of the uncertainty relation evolves as a periodic function of the acceleration factors $u_b$ and $u_c$, for which the expression has been shown in ($\ref{expressionofu}$). We note that $u$ can be viewed as a phase rotation for the modes of the fermionic field in the cavity, caused by the Bogoliubov transformation. There is no information loss in the cavity; the bound grows because some of the information encoded in entangled modes $\psi^{\text{I}}_{k_1}$, $\psi^{\text{I}}_{k_2}$, and $\psi^{\text{I}}_{k_3}$ in region~I is leaked into the other modes in region~III after acceleration. When $u=1$, the total phase rotation is an integer multiple of $2\pi$, so there is no increase of the bound resulting from acceleration.

\section{\label{sec_dis}Discussions and Conclusions}

In this work, we have addressed the question of how acceleration of quantum memories affects the fundamental limits of entropic uncertainty relations in a relativistic protocol with localized quantum fields. More precisely, to guarantee the feasibility of local quantum measurements, we have considered the protocol, which confines a spinless fermionic field inside a rigid cavity, leading to the locality of quantum information.

Our main contribution is to offer an irreducible bound for entropic uncertainty relations on a localized quantum field, and we have two principal remarks for the results presented in this paper. First, although the entropic uncertainty relation with accelerated quantum memories has been widely studied~\cite{plb-2013-fan1, plb-2015-fan2, laserpl-2017, andp-2018, epjc-2018}, that research did not clarify a relativistic protocol, revealing an uncertainty game with feasible measurements. Therefore, it would be interesting and important to understand the relations between the uncertainty principle and its variants on the localized quantum fields. To the best of our knowledge, our work is the first to bridge this connection clearly.

Second, there is a unified proof, where the data-processing inequality plays a key role, of uncertainty relations with dual entropies~\cite{prl-2012}.
However, the previous tripartite uncertainty relation only gives a state-independent lower bound, which can hardly reveal the profound connection between the acceleration of quantum memories and the uncertainties of the measured system. To overcome shortcomings of the method based on the data-processing inequality, we have employed the Holevo quantities in noninertial frames~\cite{epjc-2018}. Thus, the uncovered connection between the acceleration and the uncertainty principle not only contributes to our understanding of the essential ingredients of relativistic quantum theories, it also stimulates the intrinsic relations among quantum memory and discord~\cite{discord-2001}.

In addition to the advantages discussed above, the relativistic protocol considered here has two drawbacks. First, although our protocol is physically realizable, the potential implementation using a spin-polarized fermion trapped by an optical lattice across three cavities has not yet been experimentally investigated.
Second, to observe the effect caused by noninertial motions experimentally~\cite{Friis-prl-2013},
accelerations of agents need to be large enough to dominate the unavoidable decoherence of the system.

As a future research direction, it would be interesting to understand how a black hole with a firewall affects the bound of tripartite uncertainty relations, which also reveals the monogamy of entanglement~\cite{rmp-2017}. According to the firewall paradox of black hole, a firewall kills entanglement between an infalling particle and a late-radiated particle to guarantee entanglement monogamy, since the late-radiated particle should be entangled with early radiations~\cite{firewall-jhep-2013}. However, rigorous extension of our present work is nontrivial, thus left for future investigation.

\acknowledgments
We thank Nicolai Friis and Robert B. Mann for useful discussions and comments.
B. C. S. and C. Q. are supported by the National Natural Science Foundation of China (NSFC) with grant Grant No.~11675164.
B. C. S., Y. D. W., J. W. J. and Y. X. are supported by the Natural Sciences and Engineering Research Council of Canada (NSERC).
Y. D. W. acknowledges support from the Hong Kong Research Grant Council through grant Grant No.~17300918.
Y. X. is supported by the National Research Foundation (NRF). Singapore, under its NRFF Fellow programme (Award No. NRF-NRFF2016-02), Singapore Ministry of Education Tier 1 Grants No. MOE2017-T1-002-043 and No FQXi-RFP-1809 from the Foundational Questions Institute and Fetzer Franklin Fund (a donor-advised fund of Silicon Valley Community Foundation).  Any opinions, findings and conclusions or recommendations expressed in this material are those of the author(s) and do not reflect the views of National Research Foundation, Singapore.

\vspace{9pt}

\appendix
\section{\label{appendix_proof}Derivation of the multiple uncertainty bound~(\ref{Nmem})}

For $N+1$ quantum systems, we choose to measure $M_i\in\mathcal{M}_{n \times n}(\mathbb{C})$, where $\mathcal{M}_{n\times n}(\mathbb{C})$ is the $n\times n$
matrix space, at one subsystem~A and the rest $N$ systems are regarded as quantum memories.
The conditional entropy for the $i\text{th}$ memory $\text{E}_i$ is
\begin{equation}\label{condH}
H\left(M_i|\text{E}_i\right)=H(\rho_{M_i \text{E}_i})-H\left(\rho_{\text{E}_i}\right),
\end{equation}
where $\rho_{M_i \text{E}_i}\in\mathcal{M}_{n^2 \times n^2}(\mathbb{C})$ is the postmeasurement state of subsystems~A and $\text{E}_i$, as introduced in (\ref{eq:post-measure}). To make the derivation more clear, $\rho_{M_i \text{E}_i}$ is expanded to
\begin{equation}
\rho_{M_i \text{E}_i}=\sum^n_j p^i_j \ket{v^i_j}\bra{v^i_j} \otimes \rho^j_{\text{E}_i},
\end{equation}
where $\ket{v^i_j}$ is $j$th eigenvector of $M_i$, $p^i_j$ is the possibility to get $\ket{v^i_j}$ when we measure $M_i$ for the state of the system $\rho_\text{A}\in\mathcal{M}_{n \times n}(\mathbb{C})$, and $\rho^j_{\text{E}_i}\in\mathcal{M}_{n \times n}(\mathbb{C})$ is the final state of memory system $\text{E}_i$ when the measured state of system~A is $\ket{v^i_j}$. According to the joint entropy theorem~\cite{QI_2015} we have
\begin{equation}
H(\rho_{M_i \text{E}_i})=H(M_i)+H(\rho_{\text{E}_i|M_i}).
\end{equation}

Similar to the derivation for the bound in bipartite systems~\cite{epjc-2018}, we define a reduced density matrix $\rho_{\text{E}_i|u^i_j}\in\mathcal{M}_{n \times n}(\mathbb{C})$ which satisfies
\begin{equation}
H(\rho_{\text{E}_i|M_i})=\sum_j^n \text{Tr}(\bra{v^i_j}\rho_{M_i\text{E}_i}\ket{v^i_j}) \times H(\rho_{\text{E}_i|v^i_j}),
\end{equation}
where
\begin{equation}
\rho_{\text{E}_i|v^i_j}=\bra{v^i_j}\rho_{M_i\text{E}_i}\ket{v^i_j}/\text{Tr}(\bra{v^i_j}\rho_{M_i\text{E}_i}\ket{v^i_j}).
\end{equation}
Thus, we define a Holevo quantity
\begin{equation}
\label{Holevoq}
	\mathcal{J}(\text{E}_i|M_i)=H\left(\rho_{\text{E}_i}\right)-\sum_j^n \text{Tr}(\bra{v^i_j}\rho_{M_i\text{E}_i}\ket{v^i_j}) \times H(\rho_{\text{E}_i|v^i_j})
\end{equation}
similar to the Holevo bound~\cite{QI_2015}.
By taking $H(M_i)=\sum^n_j H(p^i_j)$ and (\ref{Holevoq}) into (\ref{condH}), we rewrite~(\ref{condH}) as
\begin{equation}
H\left(M_i|\text{E}_i\right)= \sum^n_j H(p^i_j)-\mathcal{J}(\text{E}_i|M_i).
\end{equation}

According to the derivation above,
the total entropy is
\begin{equation}
\label{Nmemeq}
\sum_i^N H\left(M_i|\text{E}_i\right) = \sum_i^N H(M_i)-\sum_i^N\mathcal{J}(\text{E}_i|M_i),
\end{equation}
with
\begin{equation}\label{SIbound}
\sum_i^N H(M_i) \geq\mathcal{B},
\end{equation}
where the lower bound~$\mathcal{B}$ is state independent and it has been introduced in Sec.~\ref{sec_approach}.
We can derive Eq.~(\ref{Nmem}) by applying~(\ref{SIbound}) to~(\ref{Nmemeq}).


\section{\label{appendix_W}Restrictions for fermion-qubit mapping in tripartite systems}
Friis \textit{et al.}~discussed the limitation of situations for mappings between a fermion and a qubit~\cite{pra-2013}. 
They argue that
the mapping is coincidentally valid for all systems with two fermionic modes when charge superselection rule is respected. However, for tripartite systems with beyond two fermionic modes, the restrictions cannot be satisfied in general cases even if the superselection rule holds. The general form of a mixed state with three modes is
\begin{align}\label{general-tri}
\rho_{k_1 k_2 k_3}= &\mu_1 \ket{0}\bra{0}+\mu_2\ket{1_{k_3}}\bra{1_{k_3}}\nonumber\\
&+\mu_3\ket{1_{k_2}}\bra{1_{k_2}}+\mu_4\ket{1_{k_2}}\ket{1_{k_3}}\bra{1_{k_3}}\bra{1_{k_2}}\nonumber\\
&+\mu_5\ket{1_{k_1}}\bra{1_{k_1}}+\mu_6\ket{1_{k_1}}\ket{1_{k_3}}\bra{1_{k_3}}\bra{1_{k_1}}\nonumber\\
&+\mu_7\ket{1_{k_1}}\ket{1_{k_2}}\bra{1_{k_2}}\bra{1_{k_1}}\nonumber\\
&+\mu_8\ket{1_{k_1}}\ket{1_{k_2}}\ket{1_{k_3}}\bra{1_{k_3}}\bra{1_{k_2}}\bra{1_{k_1}}\nonumber\\
&+(\nu_1\ket{1_{k_3}}\bra{1_{k_2}}+\nu_2\ket{1_{k_3}}\bra{1_{k_1}}\nonumber\\
&+\nu_3\ket{1_{k_2}}\bra{1_{k_1}}+\nu_4\ket{1_{k_2}}\ket{1_{k_3}}\bra{1_{k_3}}\bra{1_{k_1}}\nonumber\\
&+\nu_5\ket{1_{k_2}}\ket{1_{k_3}}\bra{1_{k_2}}\bra{1_{k_1}}\nonumber\\
    &+\nu_6\ket{1_{k_1}}\ket{1_{k_3}}\bra{1_{k_2}}\bra{1_{k_1}} + \text{H.c})\;.\nonumber\\
\end{align}
\vspace{3pt}

According to the consistency conditions to construct the reduced density matrices with two modes~\cite{pra-2013}, pairs $\{\nu_1, \nu_6\}$ and $\{\nu_3, \nu_4\}$ should have the same sign, 
whereas the pair $\{\nu_2, \nu_5\}$ should have opposite signs.

We calculate the density matrix of fermionic modes in cavities in region~III after acceleration in Sec.~\ref{sec_results}. Taking (\ref{trans from I to III}) into (\ref{W-state}), we can evaluate $\rho^3_{k_1 k_2 k_3}$ in region~III, and parameters $\{\nu_1, \nu_2, \nu_3, \nu_4, \nu_5, \nu_6\}$ are
\begin{subequations}\label{W-tri}
\begin{align}
&\nu_1=\mathcal{F}_{k_2}\mathcal{F}^*_{k_3} \;\; , \;\; \nu_6=0\, ,\\
&\nu_2=(1-f^-_{k_2})\mathcal{F}^*_{k_3} \;\; , \;\; \nu_5=-f^-_{k_2}\mathcal{F}^*_{k_3}\, ,\\
&\nu_3=\mathcal{F}^*_{k_2}(1-f^-_{k_3})\;\; , \;\; \nu_4=\mathcal{F}^*_{k_2}f^-_{k_3}\, ,
\end{align}
\end{subequations}
where $0 \leq f^-_{k_2},f^-_{k_3} \leq 1$ and $\mathcal{F}_{k_2},\mathcal{F}_{k_3}$ are expressed in (\ref{parameter}). From (\ref{W-tri}) it is easy to find that the W state is a special case where three fermionic modes can be consistently mapped to tripartite qubits without violating the restrictions.

\bibliography{paper-relativistic-uncertainty}
\end{document}